\documentclass[twocolumn,showpacs,showkeys]{revtex4}
\usepackage{graphicx}
\usepackage{bm}
\usepackage{color}
\usepackage{amsmath}
\usepackage{natbib}

\begin{document}

\title{Ion sound and dust acoustic waves at finite size of plasma particles}

\author{Pavel A. Andreev}
\email{andreevpa@physics.msu.ru}
\author{L. S. Kuz'menkov}%
\email{lsk@phys.msu.ru}
\affiliation{M. V. Lomonosov Moscow State University, Moscow, Russian Federation.}

\date{\today}

\begin{abstract}
We consider influence of finite size of ions on properties of classic plasmas. We focus our attention on the ion sound for electron-ion plasmas. We also consider dusty plasmas, where we account finite size of ions and particles of dust and consider the dispersion of dust acoustic waves. Finite size of particles affects classical plasma properties. Finite size of particles gives considerable contribution for small wave lengths, which is area of appearing of quantum effects. Consequently, it is very important to consider finite size of ions in quantum plasmas as well.
\end{abstract}


\maketitle









\section{Introduction}

Studying plasmas we assume that all particles are point-like objects. It always correct for electrons, since they have not reveal their finite size in any fundamental effects. So they are considered to be point-like even at high energy collisions reaching on accelerators. However, ions have finite radius of order of several of the Bohr radius. Considering classical gas plasma we have dial with waves having wavelength mach more interparticle distances, which is more than radius of ions. We have different situations in solids and quantum plasmas, where it is important to consider wavelengths of order of lattice constants, which have same order as size of ions. Consequently it is necessary to consider contribution of radius of ions in plasma of metals and astrophysical quantum plasmas. There are dusty plasmas containing particles of dust with sizes mach more of size of a single molecule. Consequently we have to include size of particles of dust to get correct theory of dusty plasmas.

Main goal of this paper is to account size of ions and dust particles in hydrodynamic equations and the quantum hydrodynamics as well. We present a method of finite size consideration for general hydrodynamic equations. We particularly describe consequence of finite sizes for linear waves.

As application of developed approximation we consider linear waves in electron-ion and electron-ion-dusty plasmas in absence of external fields.

Quantum plasmas reveals effects related to the quantum Bohm potential and spin dynamics \cite{MaksimovTMP 1999}-\cite{Andreev Asenjo 13}. Quantum effects change well-known properties of classic plasmas (for recent examples see Refs. \cite{Asenjo PL A 12}-\cite{Mushtaq PP EPJD 11}). Moreover, quantum plasmas open possibility for existence new effects enriching plasmas physics. As examples of new effects we can mention existence of new types of waves due to spin evolution \cite{Andreev VestnMSU 2007}, \cite{Brodin PRL 08}, \cite{Andreev IJMP 12}, appearance of instabilities caused by neutron beams propagation through magnetized plasmas due to the spin-spin and spin-current interaction of neutron spin with spins and currents in plasmas \cite{Andreev IJMP 12}, \cite{Andreev AtPhys 08}, and existence of quantum vortical structures \cite{Mahajan PL A 13}, \cite{Mahajan PRL 11}. Methods of dialing with the short-range interaction between quantum particles was developed in Ref. \cite{Andreev PRA08}.

Quantum effects and effects of finite size of particles give correction to classical effects particularly important for dense plasmas, where wave length of collective excitations can be comparable with radius of ions. Consequently quantum properties of plasmas can be under stronger influence of ion size than classical properties.

Classical properties of dusty plasmas can be affected by size of particles of dust even more than size of ions due to large average radius of dust particles. Physics of dusty (complex) plasmas reveals to be rather complicated topic, which includes different properties of dust some of them are presented in reviews \cite{Morfill RMP 09}, \cite{Fortov PReports}, \cite{Bonitz RPP 10}. Simultaneous consideration of many properties of dust makes study of waves in dusty plasmas almost impossible task. So waves have been studied in terms of simpler approaches.

Hydrodynamic and kinetic models have been used for description of collective excitations in dusty plasmas. Dust has been considered as an extra species of particles having large mass and charge in compare with the mass and charge of ions and neutrals (see for instance \cite{Prudskikh PP 13} for hydrodynamics and \cite{Gong PP 12} for kinetics). Large size of dust leads to a large collision rate between dust $\textit{and}$ ions and electrons \cite{Pandey PP 12 063}, \cite{Pandey PP 12 093}, \cite{Tolias PP 13}. It can also reveals in sticking or coming off of electrons and ions to dust giving change of charge of particles of dust. The effect of dust size distribution is considered in Refs. \cite{Benlemdjaldi PP 13}, \cite{Tripathi PP 96},
\cite{Ma PP 13}, \cite{Yang PP 13}. In this model dust is presented as set of several species with different charges $q_{dk}$, masses $m_{dk}$. In this case the Poisson equation has form
\begin{equation}\label{IS} \triangle\varphi=-4\pi\biggl(en_{i}-en_{e}+\sum_{k}q_{dk}n_{dk}\biggr).\end{equation}
Further development of the model includes that the dust grains size distribution can be considered as continuous. However, the size of dust particles is not included explicitly in this model.

The electron and ion thermal forces acting on a charged grain were discussed in the context of dusty plasmas in
Ref. \cite{Khrapak PP 13}, these forces arise due to temperature gradients in dusty plasmas.

Magnetohydrodynamics of dusty plasmas was considered by P. K. Shukla and H. U. Rahman in Ref. \cite{Shukla PP 96}. Extended hydrodynamic model including equation for the pressure evolution was also used for study of dusty plasmas \cite{Shahmansouri PP 13}. Hydrodynamics of partially ionized dusty magnetoplasmas are presented in Ref. \cite{Kopp PP 97}.

The charge fluctuation equation for the dust charge is suggested and used in literature since charge of dust particles changes during inelastic collisions and interaction with radiation  see Refs. \cite{Pandey PRE 99}, \cite{Vranjes PP 02}. An example of inelastic collisions consideration in terms of the collision integral in kinetic theory can be found in Ref. \cite{Galvo PP 12}.

Dusty plasmas reveals existence of sound waves. This topic has been studied for a long time \cite{Rao PSS 90}-\cite{Zamanian NJP 09}. Different classical effects on wave properties of dusty plasmas were reviewed by V. E. Fortov et al. in Ref. \cite{Fortov PReports}. See also recent discussion given by A. A. Mamun and P. K. Shukla \cite{Mamun JPP 11}.

Influence of a polarization force on dust acoustic waves has been considered as well \cite{Khrapak PRL 09}, \cite{Hamaguchi PP 94}, \cite{Hamaguchi PRE 94}. The polarization force $\textbf{F}_{p}$ reveals in changing of the electric force $Qn\textbf{E}=-Qn\nabla\varphi$. Considering the electric force and the polarization force together we get dressed electric force $\textbf{F}_{D}=-Qn\nabla\varphi(1-R)$, where $R=\frac{1}{4}(\mid Q\mid e/\lambda_{D}T_{i})\times$$\biggl(1-\frac{T_{e}}{T_{i}}\biggr)$, where $Q$ is the charge of dust particles, $T_{e}(T_{i})$ is the temperature of electrons (ions), $\lambda_{D}$ (see \cite{Khrapak PRL 09} and discussion in beginning of Ref. \cite{Ashrafi JPP 13}).

This paper is organized as follows. In Sec. II we show a method of introduction of finite size of ions formulating new approach. In Sec. III we consider waves in electron-ion plasmas focusing our attention on dispersion of ion-sound. In Sec. IV  we calculate dispersion of dust-sound.
In Sec. V brief summary of obtained results is presented.

\section{Model}

Using the self-consistent field-Poisson approximation for electric field
in plasma we can write hydrodynamic equations for electron-ion plasmas in the following form
\begin{equation}\label{IS continuity equation el}\partial_{t}n_{e}+\nabla (n_{e}\textbf{v}_{e})=0,
\end{equation}
\begin{equation}\label{IS Euler el} m_{e}n_{e}(\partial_{t}+\textbf{v}_{e}\nabla)\textbf{v}_{e}+\nabla p_{e}=q_{e}n_{e}\textbf{E},
\end{equation}
\begin{equation}\label{IS continuity equation ion}\partial_{t}n_{i}+\nabla (n_{i}\textbf{v}_{i})=0,
\end{equation}
and
\begin{equation}\label{IS Euler ion} m_{i}n_{i}(\partial_{t}+\textbf{v}_{i}\nabla)\textbf{v}_{i}+\nabla p_{i}=q_{i}n_{i}\textbf{E},
\end{equation}
with $m_{e}$ and $m_{i}$ ($q_{e}$, $q_{i}$) are the masses (charges) of electrons and ions, $n_{a}$ and $\textbf{v}_{a}$ are the particle concentrations and velocity fields for electrons $a=e$ and ions $a=i$, $\partial_{t}$ is the partial derivative on time, $\nabla$ is the spatial derivative (the gradient operator), $p_{a}$ is the thermal pressure, $\textbf{E}$ is the self-consistent electric field.

We have set of hydrodynamic equations (the continuity and Euler equations) for each species containing self-consistent electric field. This field is created by electrons and ions. Thus hydrodynamic equations are coupled by the quasi-electrostatic Maxwell equations
\begin{equation}\label{IS divE}\nabla \textbf{E}(\textbf{r},t)=4\pi \sum_{a}q_{a}n_{a}(\textbf{r},t),
\end{equation}
\begin{equation}\label{IS curlE}\nabla\times \textbf{E}(\textbf{r},t)=0,
\end{equation}
where (\ref{IS divE}) is the Poisson equation.

At derivation of hydrodynamic equations they appear in an integral form. Non-integral form of hydrodynamic equations is obtained at explicit introducing of the electric field created by electrons and ions
\begin{equation}\label{IS}\varphi_{a}(\textbf{r},t)=q_{a}\int G(\textbf{r},\textbf{r}') n_{a}(\textbf{r}',t) d\textbf{r}',
\end{equation}
with $\textbf{E}_{a}(\textbf{r},t)=-\nabla\varphi_{a}(\textbf{r},t)$ as usual.
Total electric field $\textbf{E}=\textbf{E}_{e}+\textbf{E}_{i}$ satisfies the Maxwell equations (\ref{IS divE}) and (\ref{IS curlE}). For consideration of particles with finite radius we need to go back to integral form of hydrodynamic equations:
the continuity equation for electrons
\begin{equation}\label{IS continuity equation el 2}\partial_{t}n_{e}+\nabla (n_{e}\textbf{v}_{e})=0,
\end{equation}
the Euler equation (the momentum balance equation) for electrons in the integral form
$$m_{e}n_{e}(\partial_{t}+\textbf{v}_{e}\nabla)\textbf{v}_{e}+\nabla p_{e}$$
$$=-q_{e}n_{e}(\textbf{r},t)\Biggl(q_{e}\nabla\int
d\textbf{r}'G(\textbf{r},\textbf{r}')n_{e}(\textbf{r}',t)$$
\begin{equation}\label{IS Euler el 2 int} +q_{i}\nabla\int
d\textbf{r}'G(\textbf{r},\textbf{r}')n_{i}(\textbf{r}',t)\Biggr),
\end{equation}
the continuity equation for ions
\begin{equation}\label{IS continuity equation ion 2}\partial_{t}n_{i}+\nabla (n_{i}\textbf{v}_{i})=0,
\end{equation}
and the Euler equation for ions in the integral form
$$m_{i}n_{i}(\partial_{t}+\textbf{v}_{i}\nabla)\textbf{v}_{i}+\nabla p_{i}$$
$$=-q_{i}n_{i}(\textbf{r},t)\Biggl(q_{e}\nabla\int
d\textbf{r}'G(\textbf{r},\textbf{r}')n_{e}(\textbf{r}',t)$$
\begin{equation}\label{IS Euler ion 2 int} +q_{i}\nabla\int
d\textbf{r}'G(\textbf{r},\textbf{r}')n_{i}(\textbf{r}',t)\Biggr),
\end{equation}
where
\begin{equation}\label{IS}G(\textbf{r},\textbf{r}')=\frac{1}{\mid\textbf{r}-\textbf{r}'\mid}
\end{equation}
is the Green function for the Coulomb interaction. Integrals in equations (\ref{IS Euler el 2 int}) and (\ref{IS Euler ion 2 int}) are over whole space. Thus a point $\textbf{r}-\textbf{r}'$ is also included. It corresponds to point like particles. For consideration of finite radius of ions we need to restrict area of integration taking integral over whole space except a sphere of radius $r_{0}=r_{i}$ for electron-ion interaction, and $r_{0}=2r_{i}$ for ion-ion interaction, where $r_{i}$ is a radius of ions.

In quantum plasmas there is additional quantum pressure: the quantum Bohm potential. Thus the left-hand side of the Euler equation contains
\begin{equation}\label{IS Bohm addition to pressure} \nabla p_{a}\rightarrow \nabla p_{a}-\frac{\hbar^{2}}{4m}\nabla\triangle
n+\frac{\hbar^{2}}{4m}\partial^{\beta}\Biggl(\frac{\nabla n\cdot\partial^{\beta}n}{n}\Biggr).\end{equation}
In the linear approximation the quantum Bohm potential gives contribution in the thermal velocity given by the temperature of the system (the Fermi temperature for degenerate plasmas).

\section{Linear waves in electron-ion plasmas}

In linear approximation on small perturbations of hydrodynamic variables $\delta n=N\exp(-i\omega t+i\textbf{k}\textbf{r})$ and $\delta \textbf{v}=\textbf{U}\exp(-i\omega t+i\textbf{k}\textbf{r})$ the set of hydrodynamic equations has following form:
the linearized form of the continuity equation for electrons
\begin{equation}\label{IS continuity equation el lin}-\imath\omega\delta n_{e}+n_{0e}\nabla\delta \textbf{v}_{e}=0,
\end{equation}
the linearized form of the Euler equation for electrons
$$-\imath\omega m_{e}n_{0e}\delta \textbf{v}_{e}+m_{e}v_{Te}^{2}\nabla\delta n_{e}$$
$$=-q_{e}n_{0e}\biggl(q_{e}\nabla\delta n_{e}\cdot\int_{0}^{+\infty}
d\textbf{r}'G(\textbf{r},\textbf{r}')e^{\imath\textbf{k}(\textbf{r}'-\textbf{r})}$$
\begin{equation}\label{IS Euler eq el lin}+q_{i}\nabla\delta n_{i}\cdot\int_{r_{0}}^{+\infty}
d\textbf{r}'G(\textbf{r},\textbf{r}')e^{\imath\textbf{k}(\textbf{r}'-\textbf{r})}\biggr),
\end{equation}
the linearized form of the continuity equation for ions
\begin{equation}\label{IS continuity equation ion lin}-\imath\omega\delta n_{i}+n_{0i}\nabla\delta \textbf{v}_{i}=0,
\end{equation}
and the linearized form of the Euler equation for ions
$$-\imath\omega m_{i}n_{0i}\delta \textbf{v}_{i}+m_{i}v_{Ti}^{2}\nabla\delta n_{i}$$
$$=-q_{i}n_{0i}\biggl(q_{e}\nabla\delta n_{e}\cdot\int_{r_{0}}^{+\infty}
d\textbf{r}'G(\textbf{r},\textbf{r}')e^{\imath\textbf{k}(\textbf{r}'-\textbf{r})}$$
\begin{equation}\label{IS Euler eq ion lin}+q_{i}\nabla\delta n_{i}\cdot\int_{2r_{0}}^{+\infty}
d\textbf{r}'G(\textbf{r},\textbf{r}')e^{\imath\textbf{k}(\textbf{r}'-\textbf{r})}\biggr),
\end{equation}
where integrals $\int_{0}^{+\infty}$, $\int_{r_{0}}^{+\infty}$, and $\int_{2r_{0}}^{+\infty}$ are integrals over whole three dimensional space except of a sphere of radius $r=0$, $r_{0}$, $2r_{0}$ correspondingly, with center at $x=y=z=0$.
This restriction of area of integration follows from the finite size of ions. The finite size of ions is illustrated with Fig. (1). Similarly we consider finite size of dust particle in the next section.

Taking integrals in the right-hand sides of the Euler equations (\ref{IS Euler eq el lin}) and (\ref{IS Euler eq ion lin}) we get that these integrals $I_{ab}$ equal to
\begin{equation}\label{IS}I_{ee}=\int_{0}^{+\infty}
d\textbf{r}'G(\textbf{r},\textbf{r}')e^{\imath\textbf{k}(\textbf{r}'-\textbf{r})}=\frac{4\pi}{k^{2}},
\end{equation}
$$I_{ei}=I_{ie}$$
\begin{equation}\label{IS}=\int_{r_{0}}^{+\infty}
d\textbf{r}'G(\textbf{r},\textbf{r}')e^{\imath\textbf{k}(\textbf{r}'-\textbf{r})}=\frac{4\pi}{k^{2}}\cos(r_{0}k),
\end{equation}
and
\begin{equation}\label{IS}I_{ii}=\int_{2r_{0}}^{+\infty}
d\textbf{r}'G(\textbf{r},\textbf{r}')e^{\imath\textbf{k}(\textbf{r}'-\textbf{r})}=\frac{4\pi}{k^{2}}\cos(2r_{0}k).
\end{equation}

Set of equations (\ref{IS continuity equation el lin})-(\ref{IS Euler eq ion lin}) can be rewritten as
\begin{equation}\label{IS disp eq short form}m_{a}(-\omega^{2}+v_{Ta}^{2}k^{2})\delta n_{a}=-q_{a}n_{0a}k^{2}\sum_{b}q_{b}I_{ab}\delta n_{b},
\end{equation}
where $a$ and $b$$=e,i$. This set can be easily generalized on system of three or more species of particles.

In our case equation (\ref{IS disp eq short form}) takes form of
$$\biggl(-{\color{blue}\omega^{2}}+v_{Te}^{2}k^{2}+\omega_{Le}^{2}\biggr)\delta n_{e}$$
\begin{equation}\label{IS final lin set for e-i_1}+\frac{4\pi q_{e}q_{i}n_{0e}}{m_{e}}\cos (r_{0}k) \delta n_{i}=0,\end{equation}
and
$$\frac{4\pi q_{e}q_{i}n_{0e}}{m_{i}}\cos (r_{0}k) \delta n_{e}$$
\begin{equation}\label{IS final lin set for e-i_2} +\biggl(-\omega^{2}+{\color{blue}v_{Ti}^{2}k^{2}}+\omega_{Li}^{2}\cos(2r_{0}k)\biggr)\delta n_{i}=0,
\end{equation}
where
$\omega_{La}^{2}=\frac{4\pi q_{a}^{2}n_{0}}{m_{a}}$
is the Langmuir frequency.

Dispersion of the ion-sound appears in the following range of frequencies
\begin{equation}\label{IS} kv_{Ti}\ll\omega\ll kv_{Te}.
\end{equation}
Using this approximation we can neglect by "blue" terms in set of equations (\ref{IS final lin set for e-i_1}) and (\ref{IS final lin set for e-i_2}) (the first term in the first group of terms in equation (\ref{IS final lin set for e-i_1}) and the second term in the second group of terms in equation (\ref{IS final lin set for e-i_2})).

In standard approach of point like ions one finds solution for the dispersion of the ion sound
\begin{equation}\label{IS} \omega^{2}=\frac{\omega_{Li}^{2}}{1+\frac{\omega_{Le}^{2}}{v_{Te}^{2}k^{2}}}=\ \biggl[\begin{array}{ccc} k^{2}v_{s}^{2}&
for &
k^{2}r_{De}^{2}\ll1\\
\omega_{Li}^{2} &
for &
k^{2}r_{De}^{2}\gg1 \\
\end{array},
\end{equation}
where $r_{De}=v_{Te}/\omega_{Le}$ is the electron Debye radius.

General form of the ion-sound dispersion including finite size of ions appears to be
\begin{equation}\label{IS ion sound R w gen} \omega^{2}=\omega_{Li}^{2}\frac{\cos(2r_{0}k)-\frac{\omega_{Le}^{2}}{v_{Te}^{2}k^{2}}\sin^{2}(r_{0}k)}{1+\frac{\omega_{Le}^{2}}{v_{Te}^{2}k^{2}}}.
\end{equation}

In "short" wave length limit $r_{e}k\gg1$ (or in other form it looks like $v_{Te}k/\omega_{Le}\gg1$)
we find dispersion of the ion-sound at account of the finite size of ions
\begin{equation}\label{IS ion sound R w Li} \omega^{2}=\omega_{Li}^{2}\cos2r_{0}k.
\end{equation}

In the large wave length limit $r_{e}k\ll1$ we obtain
\begin{equation}\label{IS ion sound R w vk} \omega^{2}=v_{s}^{2}k^{2}\biggl(\cos2r_{0}k-\frac{\omega_{Le}^{2}}{v_{Te}^{2}k^{2}}\sin^{2}r_{0}k\biggr),
\end{equation}
where $v_{s}^{2}=m_{e}v_{Te}^{2}/m_{i}$.

We have found contribution of finite size of ions in dispersion of the ion sound.
\begin{figure}
\includegraphics[width=8cm,angle=0]{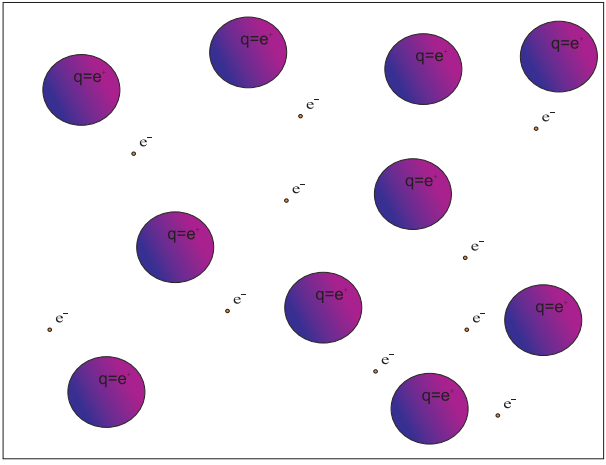}
\caption{\label{ESS 3} (Color online) The figure shows a system of electrons and ions.  The electrons are presented by small circles and ions are presented by large ones showing the finite size of ions, which is explicitly taken in the hydrodynamic equations.}
\end{figure}

\subsection{Estimation of approximation}
Paying attention to quantum plasmas let us make a note on magnitude of possible wavelengths. In classic physics the wave length of matter waves is limited by average interparticle distance. In quantum mechanics we hit the de-Broglie wave nature of particles, hence matter waves can continuously convert into collective quantum excitations with wave lengths smaller than interparticle distance.

In classical plasmas $1/k\gg r_{0}$. Consequently $\cos (r_{0}k)$ is slightly lower than 1. In this limit contribution of the quantum Bohm potential is negligible since it several orders smaller than $v_{Te}^{2}k^{2}$.

In extremely large wave vector limit, which can be reached at high density objects the contribution of the quantum Bohm potential can be comparable with the "thermal" contribution, which has to be replaced by the contribution of Fermi pressure for degenerate objects.

Getting to quantum limit we see that $1/k$ becomes smaller getting closer to $r_{0}$, so $r_{0}k$ grows up to $\pi$. As consequences we find that contribution of the Langmuir frequencies (all except the electron Langmuir frequency) can be made equal to zero or change sign.

Dispersion equation for small perturbations in quantum electron-ion plasmas allows to consider effects described above as follows
$$\Biggl(-{\color{blue}\omega^{2}}+v_{Te}^{2}k^{2}+{\color{blue}\frac{\hbar^{2}}{4m_{e}^{2}}k^{4}}+\omega_{Le}^{2}\Biggr)\delta n_{e}$$
\begin{equation}\label{IS final set foe ei quant 1} +\frac{4\pi q_{e}q_{i}n_{0e}}{m_{e}}{\color{blue}\cos (r_{0}k)} \delta n_{i}=0,\end{equation}
and
$$\frac{4\pi q_{e}q_{i}n_{0e}}{m_{i}}{\color{blue}\cos (r_{0}k)} \delta n_{e}$$
\begin{equation}\label{IS final set foe ei quant 2} +\Biggl(-\omega^{2}+{\color{blue}v_{Ti}^{2}k^{2}}+{\color{blue}\frac{\hbar^{2}}{4m_{i}^{2}}k^{4}}+\omega_{Li}^{2}{\color{blue}\cos(2r_{0}k)}\Biggr)\delta n_{i}=0.\end{equation}
Blue terms in formulas (\ref{IS final set foe ei quant 1}) and (\ref{IS final set foe ei quant 2}) are to be neglected at consideration of the classic ion-sound without account of finite size of ions. In quantum plasmas we may get to the short wave-length limit and contribution of the quantum Bohm potential becomes considerable. In this case contribution of finite size of particles is also very important and is more considerable than in classic case described above.

At $r_{0}k=\pi/2$ equations (\ref{IS final set foe ei quant 1}) and (\ref{IS final set foe ei quant 2}) become independent. Conditions for existence of nonzero perturbations $\delta n_{e}$, $\delta n_{i}$ are
\begin{equation}\label{IS} \omega^{2}=\omega_{Le}^{2}+\frac{\pi^{2}v_{Te}^{2}}{4r_{0}^{2}}+\frac{\pi^{4}\hbar^{2}}{4m_{e}^{2}}\biggl(\frac{1}{2r_{0}}\biggr)^{4}\end{equation}
for wave of electrons, and
\begin{equation}\label{IS} \omega^{2}=\frac{\pi^{2}v_{Ti}^{2}}{4r_{0}^{2}}+\frac{\pi^{4}\hbar^{2}}{4m_{i}^{2}}\biggl(\frac{1}{2r_{0}}\biggr)^{4}-\omega_{Li}^{2}\end{equation}
for waves of ions.

At large wave vectors $k\in(\frac{\pi}{4r_{0}},\frac{3\pi}{4r_{0}})$ formula (\ref{IS ion sound R w gen}) gives imaginary frequency. Formally this condition arises as
$\cos2r_{0}k-\frac{\omega_{Le}^{2}}{v_{Te}^{2}k^{2}}\sin^{2}r_{0}k\approx\cos2r_{0}k<0$. The frequency becomes real in area of larger wave vectors $k\in(\frac{3\pi}{4r_{0}},\frac{5\pi}{4r_{0}})$.

\section{Linear waves in electron-ion-dusty plasmas}
Let us present a brief description of process have been considered in quantum dusty plasmas. The QHD model was used to study different small amplitude excitations in quantum dusty magnetoplasmas paying attention to the quantum Bohm potential \cite{Jamil PP 12}, \cite{Bashir PP 12}. Low frequency non-linear waves in quantum dusty plasmas are considered in Ref. \cite{Abdelsalam PP 12} assuming that electrons and ions obey the Thomas-Fermi distribution. Hydrodynamical equations were applied to dust, but it was done with no account of the quantum Bohm potential. Solitons in quantum plasmas were considered in Ref. \cite{Wang PP 12} in terms of following model: dust is described as a classic plasma, then electrons and ions  are considered as inertialess, in compare with dust, particles described by the quantum hydrodynamics. The dust ion-acoustic solitary waves in non-planar cylindrical or spherical geometry were considered for quantum plasmas \cite{Tasnim PP 12}. Dusty plasmas of particles with internal magnetization were analyzed in Ref. \cite{Zamanian NJP 09}. However, the quantum Bohm contributions in the Euler was not included and the magnetic moment evolution was aborted.

Consideration of quantum effects in dusty plasmas is a generalization of the classic dusty plasma properties. However we are going to consider another generalization of classic dusty plasmas by consideration of the finite size of dust illustrated with Fig. (2).

In this section we consider influence of the finite size of dust particles on the dispersion of dust-acoustic waves under assumption of monosized dust particles originally derived in Ref. \cite{Rao PSS 90}. To this end we consider following set of hydrodynamic equations: the continuity equation
\begin{equation}\label{IS continuity equation dust 2}\partial_{t}n_{d}+\nabla (n_{d}\textbf{v}_{d})=0,
\end{equation}
the Euler equation (the momentum balance equation) for dust particles in the integral form
$$m_{d}n_{d}(\partial_{t}+\textbf{v}_{d}\nabla)\textbf{v}_{d}+\nabla p_{d}$$
$$=-Q n_{d}(\textbf{r},t)\Biggl(q_{e}\nabla\int
d\textbf{r}'G(\textbf{r},\textbf{r}')n_{e}(\textbf{r}',t)$$
$$+q_{i}\nabla\int
d\textbf{r}'G(\textbf{r},\textbf{r}')n_{i}(\textbf{r}',t)$$
\begin{equation}\label{IS Euler dust 2 int} +Q\nabla\int
d\textbf{r}'G(\textbf{r},\textbf{r}')n_{d}(\textbf{r}',t)\Biggr),
\end{equation}
where we have three term on the right-hand side. The first (second, last) of them presents the Coulomb interaction of dust with the electric field created by electrons (ions, dust). We consider equations for classic dusty plasmas, but we can easily come to quantum case using replacement (\ref{IS Bohm addition to pressure}).
\begin{figure}
\includegraphics[width=8cm,angle=0]{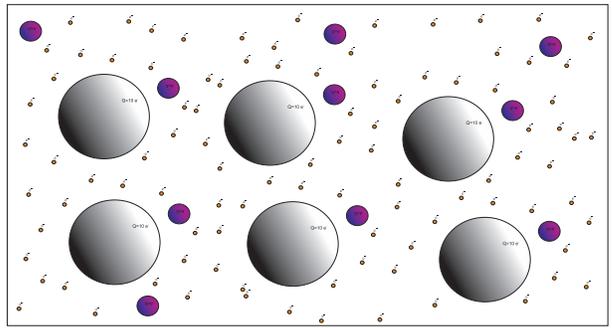}
\caption{\label{ESS 3} (Color online) The figure shows a system of electrons, ions and dust particles. We have chosen dust particles have charge Q=10e. The electrons, ions, dust particles are presented by small, intermediate, and large circles correspondingly. Size of ions is about $r_{0(ion)}=100$pm, then radius of dust particles varies in large range. For an estimation we can chose an intermediate value $r_{d}=100$nm. Figure does not show rate of radiuses of ions and dust particles. Picture presents the fact that dust particles is larger than ions.}
\end{figure}

Quasi-neutrality in equilibrium leads to $n_{0i}=n_{0e}+Zn_{0}$, where $n_{0i}$, $n_{0e}$, and $n_{0d}$ are the equilibrium particle concentration of ions, electrons and dust. $Z$ is the charge number of dust particles $Q/e$. Considering plane wave perturbations of the concentration and velocity field of dust we get linearized set of hydrodynamic equations
\begin{equation}\label{IS cont eq conseq} \textbf{k}\delta \textbf{v}= \frac{\omega\delta n}{n_{0}},\end{equation}
and
$$ -\imath\omega m_{d}n_{0d}\delta \textbf{v}_{d}+m_{d}v_{Td}^{2}\nabla \delta n_{d}$$
$$=-Qn_{0d}\Biggl(q_{e}\nabla\int_{"r_{d}"}
d\textbf{r}'G(\textbf{r},\textbf{r}')n_{e}(\textbf{r}',t)$$
$$+q_{i}\nabla\int_{"r_{d}+r_{i}\approx r_{d}"}
d\textbf{r}'G(\textbf{r},\textbf{r}')n_{i}(\textbf{r}',t)$$
\begin{equation}\label{IS Euler eq conseq} +Q\nabla\int_{"2r_{d}"}
d\textbf{r}'G(\textbf{r},\textbf{r}')n_{d}(\textbf{r}',t)\Biggr).\end{equation}

We assume that the densities
of electrons and ions satisfy the Boltzmann distribution $n_{a}=n_{0a}\exp(-q_{a}\varphi/T_{a})$, where $\varphi$ is the potential of field created by perturbations of dust $\varphi=-Q\int_{"2r_{d}"}
d\textbf{r}'G(\textbf{r},\textbf{r}')n_{d}(\textbf{r}',t)$.

We find solution in assumption of high temperature of electrons and ions: $n_{a}=n_{0a}(1-q_{a}\varphi/T_{a})$.
In this case the set of equations (\ref{IS cont eq conseq}) and (\ref{IS Euler eq conseq}) can be represent as
$$-\imath m_{d}(\omega^{2}-k^{2}v_{Td}^{2})\delta n_{d}$$
$$ =-Qn_{0}\biggl(Q\textbf{k}\nabla\int G\delta n_{d}d\textbf{r}'$$
\begin{equation}\label{IS Euler eq lin high temp with int}+e^{2}\biggl(\frac{n_{0e}}{T_{e}}+\frac{n_{0i}}{T_{i}}\biggr)\textbf{k}\nabla\int G\varphi d\textbf{r}'\biggr),\end{equation}
where we included quasi-neutrality condition $\int G(\textbf{r},\textbf{r}')(-en_{0e}+en_{0i}+Qn_{0d})d\textbf{r}'=0$.

Taking integrals we get final form of dispersion equation
$$ m_{d}(\omega^{2}-k^{2}v_{Td}^{2})\delta n_{d}=$$
$$-Q^{2}n_{0}k\imath\cos(kr_{d})\biggl[\frac{4\pi}{k^{2}}\textbf{k}\nabla\delta n_{d}$$
\begin{equation}\label{IS Euler eq lin high temp with int are taken} -e^{2}\biggl(\frac{4\pi}{k^{2}}\biggr)^{2}\cos(2kr_{d})\biggl(\frac{n_{0e}}{T_{e}}+\frac{n_{0i}}{T_{i}}\biggr)\textbf{k}\nabla\delta n_{d}\biggr] \end{equation}
giving following solution for frequency of waves in dusty plasmas
$$\omega^{2}=v_{Td}^{2}k^{2}+\frac{4\pi Q^{2}n_{0}}{m_{d}}\cos(kr_{d})\times$$
\begin{equation}\label{IS disp gen} \times\Biggl[1-\frac{4\pi e^{2}}{k^{2}}\cos(2kr_{d})\biggl(\frac{n_{0e}}{T_{e}}+\frac{n_{0i}}{T_{i}}\biggr)\Biggr].\end{equation}
Neglecting the first term for cold dust we can rewrite formula (\ref{IS disp gen}) in the traditional for the dust acoustic wave form, which was obtained for the first time in Ref. \cite{Rao PSS 90} for point-like particles,
\begin{equation}\label{IS dust-acustic waves} \omega^{2}=\beta^{2}C_{s}^{2}k^{2}\cos(kr_{d})\biggl[1+\frac{\lambda_{De}^{2}k^{2}}{(1+\eta\delta)\cos(2kr_{d})}\biggr]\end{equation}
at $\frac{\lambda_{De}^{2}k^{2}}{1+\eta\delta}\gg1$ and $kr_{d}\ll\pi/2$. In formula (\ref{IS dust-acustic waves}) we have use next designations $\beta=Z(\delta-1)/(1+\eta\delta)$, $C_{s}=\sqrt{T_{e}/m_{d}}$, $\lambda_{De}=\sqrt{T_{e}/(4\pi e^{2}n_{0})}$, $\eta=T_{e}/T_{i}$, $\delta=n_{0i}/n_{0e}$.

Some interesting consequences of finite size of particle in context of wave dispersion in dusty plasmas can be found from formula (\ref{IS disp gen}), when we do not apply conditions $\frac{\lambda_{De}^{2}k^{2}}{1+\eta\delta}\gg1$ and $kr_{d}\ll\pi/2$.
Size of dust particles is more than size of ions so we can get $kr_{0}\in(\pi/4,\pi)$ in classic regime. So we can consider consequences revealing in damping of the dust acoustic waves.

At $(1+\delta\eta)/(\lambda^{2}k^{2})<1$ we have that the square brackets in formula (\ref{IS disp gen}) is positive. And the sign of the second term in the formula is determined by $\cos(kr_{d})$. So this term becomes negative at $kr_{d}\in(\frac{\pi}{2},\frac{3\pi}{2})$. Let us also consider behaviour of terms within the square brackets. Presence of $\cos(2kr_{d})$ gives decreasing of the second term in square brackets with increasing of $kr_{0}\in(0,\pi/4)$, and it changes sign at $kr_{0}=\pi/4$ being negative at $kr_{0}\in(\pi/4,3\pi/4)$ and increasing its module at $kr_{0}\in(\pi/4,\pi/2)$. At $kr_{0}\in(\pi/2,3\pi/4)$ the second term is negative and its module decreased in compare with zero-size particles. Finally we see that at $kr_{0}\in(\pi/2,3\pi/4)$ both the contribution of $\cos(kr_{d})$ and the contribution of square brackets become negative. Hence damping of dust acoustic waves may not appear.

At $\xi\equiv(1+\delta\eta)/(\lambda^{2}k^{2})>1$ behaviour of $\cos(2kr_{d})$ gives dramatic contribution of the square brackets. This case allows to reach the following identity $1-\xi\cos(2kr_{d})=0$. That makes the second term in formula (\ref{IS disp gen}) equals to zero, so frequency appears as $\omega^{2}=v_{Td}^{2}k^{2}_{0}$, where $k_{0}$ is given by an transcendental equation $k_{0}=\frac{1}{2r_{d}}\arccos(1/\xi(k_{0}))$. Going further we may admit that $\cos(2kr_{d})>0$ at $2kr_{d}\in[0,\pi/4)\cup(3\pi/4,5\pi/4)$. However it is not enough to reach $1-\xi\cos(2kr_{d})<0$. To this end we have to consider $kr_{d}\in[0,k_{0}r_{d})\cup(\pi-k_{0}r_{d},\pi+k_{0}r_{d})$. We have two consequences of $1-\xi\cos(2kr_{d})<0$ for each of two intervals presented above. At small $k$ when $\cos(kr_{d})>0$ we get instability of the spectrum (\ref{IS disp gen}). In area $kr_{d}\in(\pi-k_{0}r_{d},\pi+k_{0}r_{d})$ when $\cos(kr_{d})<0$, in opposite, we find stabilization of instability presented above for $\xi<1$.

We can also consider now limit case of spectrum (\ref{IS dust-acustic waves}) at small $k$ applying $\cos(kr_{d})\approx1-kr_{d}/2$ that gives approximate dispersion dependence of the dust acoustic waves for finite size particles
$$\omega^{2}=\beta^{2}C_{s}^{2}k^{2}\biggl(1-\frac{(kr_{d})^{2}}{2}\biggr)\times$$
\begin{equation}\label{IS dust-acustic waves expansion} \times\biggl[1+\frac{\lambda_{De}^{2}k^{2}}{(1+\eta\delta)}\biggl(1+2(kr_{d})^{2}\biggr)\biggr].\end{equation}

The traditional dust acoustic solution does not exists at $\cos(kr_{0})\approx0$ and $\cos(2kr_{0})\approx0$. In the first case the right-hand side of equation (\ref{IS Euler eq lin high temp with int are taken}) equals to zero and we get $\omega=kv_{Td}=\frac{\pi v_{Td}}{2r_{0}}$. In the second of these cases the second term in the second group of terms of formula (\ref{IS disp gen}) equals to zero and we have $\omega^{2}=\frac{\pi^{2}v_{Td}^{2}}{16r_{0}^{2}}+\frac{2\sqrt{2}\pi Q^{2}n_{0}}{m_{d}}$ at wave vectors $k\simeq \frac{\pi}{4r_{0}}$.

\section{Conclusion}
We have presented a mechanism changing the electric force in electron-ion plasmas due to finite size of ions. This mechanism is also important for dusty plasmas, where it gives contribution in the force density along with the polarization force. We have considered contribution of the finite size of dust particles for monosized dust particles. We have illustrated contribution of the finite size of particles in the dispersion of low-frequency waves. We have also shown that the finite size of particles affects properties of quantum plasmas.

Presented method consideration of size of ions and dust particles opens possibilities for consideration of various linear and non-linear effects in different classic and quantum plasmas. This method can also be applied for kinetic equation as well.

\end{document}